\documentclass[12pt]{article}
\usepackage{latexsym}
\usepackage{amsmath}
\usepackage{amsfonts}
\usepackage{amssymb}
\usepackage{graphicx}
\usepackage{slashbox}
\usepackage{color}
\usepackage{tikz}
\usepgflibrary{arrows}

\title{Dynamics on the cone: \\closed orbits and superintegrability}
\author{
Y. Brihaye\\
\small Department of Theoretical and Mathematical Physics,\\
\small University of Mons,\\
\small 20, Place du Parc, B7000 Mons, Belgium
\and
 P. Kosi\'nski, P. Ma\'slanka\thanks{e-mail: pmaslan@uni.lodz.pl}\\
\small Department of Theoretical Physics and Computer Science, \\
\small University of \L\'od\'z,\\
\small Pomorska 149/153, 90-236 {\L}\'od\'z, Poland
}

\date{}
\begin{document}
\maketitle
\begin{abstract}
The generalization of Bertrand's theorem to the case of the motion of point particle on the surface of a cone is presented. The superintegrability of such models is discussed. The additional integrals of motion are analyzed for the case of Kepler and harmonic oscillator potentials.
\end{abstract}
\section{Introduction}
\par
The cone geometry appears in a number of physical contexts such as $2+1$-dimensional counterpart of Schwardschild solution \cite{b1}, cosmic strings \cite{b2}, defects in condensed matter physics \cite{b3}, \cite{b4} and other. 
\par
In view of possible applications the motion of a particle on the surface of a cone has been studied both classicaly and quantum mechanically \cite{b5}- \cite{b16}. In particular, the motion under the influence of central potential (i.e. depending on the distance from the tip of a cone) has attracted some attention. In a very interesting paper Al-Hashimi and Wiese \cite{b17} considered the case of Kepler and harmonic oscillator potentials. They showed that all bounded orbits are then closed provided the scale factor $s$, defined bellow by eq. (\ref{e2}), is rational. Moreover, they constructed the generators of accidental symmetries which appeared to be the straightforward generalizations of the standard (i.e. corresponding to s=1) ones. The situation becomes even more interesting in the quantum domain. One can still define the accidental symmetry generators obeying proper commutation rules. However, since one is dealing with unbounded operators, the problem of proper definition of their domains becomes urgent. It appears that, contrary to the standard case, for general $s$ being a rational multiple of $\pi $, the degenerate eigenstates of the Hamiltonian do not form the complete multiplets of symmetry algebra. 
\par
The present paper is devoted to the study of symmetry aspects of classical motion of a particle on a cone under the influence of a central potential. We show first that the Bertrand's theorem \cite{b18} generalizes to the case of particle motion on the surface of a cone. Namely, the following statement holds true:\\
{\em All bounded trajectories of a nonrelativistic particle moving on the surface of the cone under the influence of a central potential $V(r)$ are closed if and only if $V(r)=-\frac{\kappa }{r}$ or $V(r)=\frac{m\omega ^2r^2}{2}$ and the scale factor $s$ is rational.}\\
Noting further that our system:$(i)$ is integrable because there are two globally defined integrals of motion, $H$ and $J$; $(ii)$ can be obtained from the special case $s=1$ by a (locally defined) cannonical transformation, we show that the property that all bounded orbits are closed is equivalent to the superintegrability. Moreover, the additional integral of motion can be found by applying the above mentioned canonical transformation to the integral of motion for the $s=1$ case. The resulting expression is, for $s$ noninteger, defined only locally. However, for $s$ rational one immediately finds the global counterpart by taking an appropriate function of it. 
\par
Viewing our system as an example of twodimensional integrable dynamics we find that all conclusions are next to obvious.

\section{Dynamics on the cone} 
Consider the nonrelativistic particle of mass $m$ confined to the surface of a cone and bound to its tip by a potential $V(r),\; r$\
 being the distance from the tip. Let $\delta $\ be the deficit angle and $\chi $ – the polar angle (cf. Fig. 1).\\

 \begin{figure} [h]
\begin{tikzpicture}[>=stealth]
\draw (40:4cm) arc (40:320:4cm);
\draw (0mm,0mm) -- (40:4cm);
\draw (0mm,0mm) -- (320:4cm);
\draw[<-] (320:2.5cm) arc (320:360:2.5cm);
\draw[->] (0:2.5cm) arc (0:40:2.5cm);
\draw[->] (0mm,0mm) -- (70:3.5cm);
\draw[<->] (40:2.5) arc (40:70:2.5);
\draw (70:2.5cm) node[anchor=east] {$r$};
\draw (1.25cm,0cm) node {$\delta$};
\draw (55:1.25cm) node {$\chi$};
\end{tikzpicture}
\end{figure}
\begin {center}
Fig. 1
\end{center}
Then
\begin{equation}
\label{e1}
0\leq \chi <2\pi -\delta 
\end{equation}
Let us rescale the polar angle $\chi $\ as to extend it to the range $<0, 2 \pi )$:

\begin{equation}
\label{e2}
\begin{split}
&\varphi \equiv \frac{\chi }{s},\;\;\;\;0\leq \varphi <2\pi \\
&s\equiv 1-\frac{\delta }{2\pi }
\end{split}
\end{equation}
The scaling parameter $s$\ is related to the angle $\alpha $ between the symmetry axis and the generating of the cone (see Fig. 2) by the formula
\begin{equation}
\label{e3}
\begin{split}
s=sin\alpha 
\end{split}
\end{equation}

\begin{figure} [h]
\begin{tikzpicture}
\draw[->] (0cm,0cm) -- (6cm,0cm);
\draw[->] (0cm,0cm) -- (0cm,6cm);
\draw[->] (0cm,0cm) -- (-3.5cm,-3.5cm);
\draw (0cm,0cm)--(2cm,4cm);
\draw (0cm,0cm)--(-2cm,4cm);
\draw (-2cm,4cm).. controls (0cm,4.7cm) ..(2cm,4cm);
\draw (-2cm,4cm).. controls (0cm,3.3cm) ..(2cm,4cm);
\draw [<->] (-0.5cm,1cm)..controls (-0.25cm,1.25cm)..(0cm,1.25cm);
\draw (-3.25cm,-3.75cm) node {$q_1 $};
\draw (6cm,-0.30cm) node {$q_2 $};
\draw (-0.30cm,6cm) node {$q_3 $};
\draw (-0.2cm,0.80cm) node {$\alpha $};
\end{tikzpicture}
\end{figure}
\begin {center}
Fig. 2
\end{center}

The relevant Lagrangian describing the dynamics of our particle reads

\begin{equation}
\label{e4}
L=\frac{m}{2} \dot{\vec{q} } ^2-V(r),\;\;\;\;\;r=\mid \vec{q} \mid 
\end{equation}
Using
\begin{equation}
\label{e5}
 \dot{\vec{q} } ^2=\dot{r}^2+r^2\dot{\chi }^2=\dot{r}^2+r^2s^2\dot{\varphi }^2     
\end{equation}
one can write
\begin{equation}
\label{e6}
\begin{split}
L=\frac{m}{2}\dot{r}^2+\frac{mr^2s^2\dot{\varphi }^2}{2}-V(r)
\end{split}
\end{equation}
It is convenient to introduce the "Cartesian" coordinates parametrizing the plane
\begin{equation}
\label{e7}
\begin{split}
&x_1=rcos\varphi  \\
&x_2=rsin\varphi 
\end{split}
\end{equation}
Then $L$\ takes the form
\begin{equation}
\label{e8}
\begin{split}
&L=\frac{m\dot{\vec{x} }^2}{2}+\frac{m(s^2-1)}{2r^2}(\vec{x} \times \dot{\vec{x} })^2-V(r)=    \\
&=\frac{ms^2\dot{\vec{x} }^2}{2}-\frac{ m(s^2-1)}{2r^2}(\vec{x} \dot{\vec{x} })^2-V(r) 
\end{split}
\end{equation}
Moreover
\begin{equation}
\label{e9}
\begin{split}
\vec{p}\equiv \frac{\partial L}{\partial \dot{\vec{x} } }=ms^2\dot{\vec{x} }-\frac{m(s^2-1)}{r^2}(\vec{x} \dot{\vec{x} } )\vec{x}    .
\end{split}
\end{equation} 
or
\begin{equation}
\label{e10}
\dot{\vec{x} } =\frac{\vec{p}}{ms^2}+\frac{(s^2-1)}{ms^2 r^2}(\vec{x}  \vec{p})\vec{x}   .
\end{equation} 
The Hamiltonian reads
\begin{equation}
\label{e11}
\begin{split}
H=\vec{p}\dot{\vec{x} } -L=\frac{\vec{p}^2 }{2m}+\frac{1-s^2}{2ms^2r^2}J^2+V(r),
\end{split}
\end{equation} 
where $J\equiv \mid \vec{x} \times \vec{p}\mid =mr^2s^2\dot{\varphi } $\ is the angular momentum. 
Defining
\begin{equation}
\label{e12}
\begin{split}
p_r=\frac{\vec{x}\vec{p}}{r}   .
\end{split}
\end{equation} 
we find finally
\begin{equation}
\label{e13}
\begin{split}
H=\frac{p_r^2}{2m}+\frac{J^2}{2ms^2r^2}+V(r)  .
\end{split}
\end{equation} 
\par
The same results can be obtained by adding to the unconstrained Lagrangian in $q$-space the constraint multiplied by Lagrange multiplier 
(this yields what is called in classical mechanics the Lagrangian equations of the first kind) and applying the Dirac analysis of constrained systems.

\section{Bertrand's theorem and superintegrability} 
\par
We shall show that the analogue of Bertrand's theorem holds in the form presented in Sec. 1. Namely, all bounded orbits are closed only provided $V(r)$ corresponds to the Kepler problem
\begin{equation}
\label{e14}
V(r)=-\frac{\kappa }{r}
\end{equation}
or harmonic oscillator
\begin{equation}
\label{e15}
V(r)=\frac{\beta r^2}{2}
\end{equation}
and $s$\ is rational. The proof goes, with small modifications, along the same lines as in the standard case \cite{b19}.
 We shall assume that either $V(r)\rightarrow  \infty $\ or $V(r)\rightarrow  c$\ as $r\rightarrow  \infty $; in the latter case one can safely put c =0.
 Using the fact that J and E are constant of motion we easily find the equation determining the trajectory
\begin{equation}
\label{e16}
\varphi =\pm \int \frac{\frac{J}{ms^2r^2} dr}{\sqrt{\frac{2}{m}(E-\frac{J^2}{2ms^2r^2}-V(r))}}
\end{equation}
In particular, denoting by $\triangle \varphi $\ the angle between the directions of adjacent perigeum and apogeum we get
\begin{equation}
\label{e17}
s\triangle \varphi = \int_{r_{min}}^{r_{max}} \frac{\frac{\lambda}{mr^2} dr}{\sqrt{\frac{2}{m}(E-\frac{\lambda ^2}{2mr^2}-V(r))}}
\end{equation}
where $\lambda \equiv \frac{J}{s}$. The right-hand side of eq. (\ref{e17}) has the same form as in the case of plane (i.e. $s=1$) motion. Therefore, one can follow the same line of reasoning as in the standard case. In order to have a closed orbit $\frac{\Delta \varphi}{\pi} $ must be rational. Now, the right hand side of eq. (\ref{e17}) is (at least in a certain domain) a continuous function of $\lambda $ and $ E $. Therefore, it must be a constant. Making the change of variables $x=\frac {\lambda }{mr}$ we finally conclude that the integral
\begin{equation}
\label{e18}
\int^{x_{max}}_{x_{min}} \frac {dx}{\sqrt{\frac{2}{m}\Big(E-\frac{mx^2}{2}-V(\frac{\lambda }{mx})\Big)}}
\end{equation}
must be independent on $E$ and $\lambda $. It represents (half of) the period of \linebreak onedimensional motion in the potential
\begin{equation}
\label{e19}
U(x;\lambda )\equiv \frac{mx^2}{2}+V\bigg(\frac{\lambda }{mx}\bigg)
\end{equation}
It is easy to see that, for the period to be energy independent, $U(x;\lambda )$ must have unique minimum $x_0$ for any fixed $\lambda $. Now, assume $E$ to be close to the minimum value of $U(x;\lambda )$. Then one can use the harmonic oscillator approximation. Within this approximation, 
\begin{equation}
\label{e20}
\begin {split}
&\frac {T}{2}=\frac{\pi }{\omega  }\\
&\omega ^2=\frac{1}{m}\frac{d^2U(x;\lambda )}{dx^2}\mid  _{x=x_{0}}
\end {split}
\end{equation}
Eqs. (\ref{e19}) -  (\ref{e20}) imply
\begin{equation}
\label{e21}
\begin {split}
&mx_0-\frac{\lambda }{mx^2_0}V'\Big(\frac{\lambda }{mx_0}\Big)=0 \\
&\omega  ^2=1+\frac{2\lambda }{m^2x^3_0}V'\Big(\frac{\lambda }{mx_0}\Big)+\frac{\lambda ^2}{m^3x^4_0}V''\Big(\frac{\lambda }{mx_0}\Big)=0
\end {split}
\end{equation}
or
\begin{equation}
\label{e22}
\omega ^2=3+\frac{\lambda }{mx_0}\frac{V''(\frac{\lambda }{mx_0})}{V'(\frac{\lambda }{mx_0})}
\end{equation}
As it has been mentioned above $x_0$ is a unique function of $\lambda $; moreover, first eq. (\ref{e21}) implies that $\frac{\lambda} {x_0}$ varies as $\lambda $ varies. Therefore, the independence of the integral (\ref{e18}) on $E$ and $\lambda $ leads to the equation
\begin{equation}
\label{e23}
\frac {rV''(r)}{V'(r)}=C-3,  \;\ \:\ \:\ C\equiv \omega ^2>0
\end{equation}
and either
\begin{equation}
\label{e24}
V(r)=Ar^\alpha , \;\ \;\ \;\ \alpha  >-2
\end{equation}
or
\begin{equation}
\label{e25}
V(r)=Bln\Big(\frac{r}{r_0}\Big)
\end{equation}
Now, in oder to select the potentials $U(x;\lambda )$ leading to energy independent period beyond the harmonic oscillator approximation one can use the method described in Ref. \cite{b20}. Namely, $U(x;\lambda )$ should obey
\begin{equation}
\label{e26}
x_2(U)-x_1(U)=a\sqrt{U-U_0}
\end{equation}
where $a$ is some constant and the notation is explained on Fig. 3.

\begin{figure} [h]
\begin{tikzpicture}
\draw[->] (-0.5cm,0cm) -- (6cm,0cm);
\draw[->] (0cm,-0.5cm) -- (0cm,8cm);
\draw (0.5cm,7.25cm) parabola bend (3cm,1cm) (5.5cm,7.25cm);
\draw[dashed,very  thin] (0cm,1cm) -- (3cm,1cm);
\draw[dashed,very  thin] (1cm,5cm) -- (1cm,0cm);
\draw[dashed,very  thin] (0cm,5cm) -- (5cm,5cm) -- (5cm,0cm);
\draw (0cm,8cm) node[anchor=east] {$U$};
\draw (0cm,5cm) node[anchor=east] {$U$};
\draw (0cm,1cm) node[anchor=east] {$U_0$};
\draw (1cm,0cm) node[anchor=north] {$X_1(U)$};
\draw (5cm,0cm) node[anchor=north] {$X_2(U)$};
\end{tikzpicture}
\end{figure}
\begin {center}
Fig. 3
\end{center}

It is not difficult to check that the condition (\ref{e26}) excludes (\ref{e25}) and implies $\alpha =-1$ or $\alpha =2$ in the case (\ref{e24}).\\
Finally, let us compute the righ-hand side of eq. (\ref{e17}); for $\alpha =2$ we get $\frac {\pi }{2}$ while $\alpha = -1$ yields $\pi $. Going back to the eq. (\ref{e17}) we see that $s$ must be rational which concludes the proof.\\
Let us now discuss the notion of superintegrability. Assume that the Hamiltonian admits bounded motion in some region of phase space. Our system is integrable because it admits two Poisson commuting integrals of motion, $H$ and $J$. Using standard methods one can construct the Liouville-Arnold tori and the corresponding action-angle variables; the Hamiltonian is a function of the action variables $I_1$ and $I_2$ only
\begin{equation}
\label{e27}
H=H(I_1, I_2)
\end{equation}
Let $\omega _k(I)\equiv \frac{\partial H}{\partial I_k},\;\ k=1,2,$ be the frequencies corresponding to the angle variables $\varphi _k$. It is well known that the trajectories corresponding to the fixed values of $H$ and $J$ (i.e. lying on some Liouville-Arnold torus) are closed if and only if $\frac {\omega _1(I)}{\omega_2 (I)}$ is a rational number. If all bounded orbits (i.e. lying on all $L-A$ tori) are closed then, as can be easily shown, the frequencies have the form $\omega _k(I)=n_k\omega (I), \;\ k =1, 2$ where $n_k$ are some integers. In such a case there exists simple way to construct third integral of motion \cite{b19}-\cite{b21} (for a recent review of superintegrability see \cite {b22}). Namely, the above form of frequencies implies
\begin{equation}
\label{e28}
H (\underline{I})=H(n_1I_1+n_2I_2)
\end{equation}
Then $n_2\varphi _1-n_1\varphi _2$ is an integral of motion and, due to the fact that $n_{1,2}$ are integers, any trigonometric function of it is globally well-defined. So we conclude that the property of all bounded orbits being closed implies superintegrability.\\
On the other hand, if the ratio $\frac {\omega _1(I)}{\omega _2(I)}$ is an irrational number the trajectories cover densely a given torus. Therefore, they cannot result from intersection of $L-A$ torus with the level surface of the third integral of motion; the system is not superintegrable. \\
Let us point out that the conclusion concerning nonsuperintegrability is strongly related to the existence of bounded orbits. If all orbits are unbounded the invariant hypersurfaces in phase space obtained by fixing the values of commuting integrals of motion are no longer tori. We are no longer dealing with angle variables. However, it is the $2\pi $-periodicity which plays a~decisive role in the above considerations. It sets the normalization of angle variables and, consequently, the frequencies which makes it sensible to pose the question concerning their ratio. Therefore, the superintegrability is much less restrictive condition in this case.

\section{The Kepler problem} 
As we have discussed in the previous section superintegrability is equivalent to the property that all bounded orbits are closed. Therefore, in the attractive case central motion on the surface of a cone is superintegrable if and only if the potential describes the Kepler system or harmonic oscillator and $s$ is rational.\\
In this section we consider the Kepler problem. The Hamiltonian reads
\begin{equation}
\label{e29}
H=\frac{p^2r}{2m}+\frac{J^2}{2ms^2r^2}-\frac{\kappa }{r}
\end{equation}
For $s$ rational all orbits are closed so the system is superintegrable. The action variables read $($for $H<0)$
\begin{equation}
\label{e30}
I_1=\frac{1}{2\pi }\oint Jd\varphi =J
\end{equation}
\begin{equation}
\label{e31}
I_2=\frac{1}{2\pi }\oint p_rdr=\frac{1}{\pi }\int^{r_{max}}_{r_{min}}\sqrt{2m\Big(H-\frac{J^2}{2ms^2r^2}+\frac{\kappa }{r}\Big)} \quad dr
\end{equation}
This yields
\begin{equation}
\label{e32}
H=-\frac{m\kappa ^2}{2\Big(\frac{I_1}{s}+I_2\Big)^2}
\end{equation}
For rational $s$, $s=\frac{n_2}{n_1}$ one finds
\begin{equation}
\label{e33}
H=-\frac{mn^2_2\kappa ^2}{2(n_1I_1+n_2I_2)^2}
\end{equation}
in accordance with eq. (\ref{e28}) of previous section. One can now construct the~third integral of motion following the method presented above. However, the~derivation is simplified by noting that the (local) canonical transformation $J\rightarrow sJ$, $\varphi \rightarrow \frac{\varphi }{s}$ reduces the problem to the~standard two-dimensional Kepler one. Therefore, we can obtain our integral from the~corresponding Runge-Lenz one. This results in the following expression
\begin{equation}
\label{e34}
C\equiv (A-iB)e^{is\varphi }
\end{equation}
\begin{equation}
\label{e35}
A=\frac{J^2}{ms^2r}-\kappa 
\end{equation}
\begin{equation}
\label{e36}
B=\frac{Jpr}{ms}
\end{equation}
One easily checks that
\begin{equation}
\label{e37}
A^2+B^2=\frac{2HJ^2}{ms^2}+\kappa ^2
\end{equation}
$C$ is not single-valued unless $s$ is integer. For $s$ irrational nothing can be done. On the other hand, for
\begin{equation}
\label{e38}
s=\frac{k}{n}
\end{equation}
we find that
\begin{equation}
\label{e39}
Z\equiv C^n=(A-iB)^ne^{ik\varphi }
\end{equation}
is a single valued integral of motion. Now $H, J, Z$ and $\bar {Z}$ form a finite \mbox{$W$-algebra} \cite{b23}  with respect to the Poisson bracket. The nontrivial brackets read
\begin{equation}
\label{e40}
\{J, Z\}=kZ,\qquad \{J, \bar Z\}=-k\bar Z
\end{equation}
\begin{equation}
\label{e41}
\{Z, \bar Z\}=\frac{4in^3}{mk}JH\Big(\frac{2n^2J^2}{mk^2}+\kappa ^2\Big)^{n-1}
\end{equation}
We see that only for $n\equiv 1$, i.e. integer $s$, the above algebra becomes a Lie algebra on the submanifold of constant energy.

\section {Harmonic oscillator}
The relevant Hamiltonian has the form
\begin{equation}
\label{e42}
H=\frac{p^2r}{2m}+\frac{J^2}{2ms^2r^2}+\frac{m\omega ^2r^2}{2}
\end{equation}
All orbits are bounded and closed if $s$ is rational. Therefore, the system is then superintegrable. The action variables read now
\begin{equation}
\label{e43}
I_1=J
\end{equation}
\begin{equation}
\label{e44}
I_2=\frac{1}{\pi }\int^{r_{max}}_{r_{min}}\sqrt{2m\Big(H-\frac{J^2}{2m^2s^2r^2}-\frac{m\omega ^2r^2}{2}\Big)}\quad dr
\end{equation}
which leads to
\begin{equation}
\label{e45}
H= \omega \Big(\frac{I_1}{s}+2I_2\Big)
\end{equation}
For rational $s$, $s=\frac{n_2}{n_1}$, one obtains
\begin{equation}
\label{e46}
H = \frac {\omega }{n_2}\Big (n_1I_1+2n_2I_2\Big)
\end{equation}
Again, the general method discussed previously allows us to construct the third integral of motion. As in the case of Kepler problem it is sufficient to know the form of the integral in the standard case $(s=1)$ and to apply the canonical transformation $J\rightarrow sJ, \varphi \rightarrow \frac{\varphi }{s}$ which yields the locally defined constant of motion.
\begin{equation}
\label{e47}
C=(A-iB)e^{2is\varphi }
\end{equation}
\begin{equation}
\label{e48}
A=\frac{J^2}{ms^2r^2}-H
\end{equation}
\begin{equation}
\label{e49}
B=\frac{prJ}{msr}
\end{equation}
where 
\begin{equation}
\label{e50}
A^2+B^2=H^2-\frac{\omega ^2J^2}{s^2}
\end{equation}
Now, for rational $s=\frac{k}{n}$ one constructs the globally defined integral of motion by
\begin{equation}
\label{e51}
Z=C^n=(A-iB)^ne^{2ik\varphi }
\end{equation}
The integrals $Z, \bar {Z}, J$ and $H$ generate a finite $W$-algebra described by the following nontrivial Poisson brackets
\begin{equation}
\label{e52}
\{J, Z\}=2kZ
\end{equation}
\begin{equation}
\label{e53}
\{J, \bar  {Z}\}=-2k\bar {Z}
\end{equation}
\begin{equation}
\label{e53}
\{Z,  \bar  {Z}\}=-\frac{4in^3}{k}\omega ^2J\Big(H^2-\frac{\omega ^2n^2J^2}{k^2}\Big)^{n-1}
\end{equation}
For $n=1$, i.e. integer $s$, we arrive at the Lie algebra.

\section {Conclusions}
\par
We considered the motion of a particle on a cone under the influence of central potential $V(r)$. Such a system is integrable since there are two commuting integrals of motion, $H$ and $J$. Assume that $V(r)$ is such that there are bounded orbits. We have shown that all such orbits are closed if and only if $V(r)=-\frac{\kappa }{r}$ or $V(r)=m\omega ^2\frac{r^2}{2}$ \underline {and} $s$ is rational. By the same arguments as in the standard $(s=1)$ case the dynamics is then superintegrable (and only then provided there are bounded orbits).
\par
Our proof can be reformulated as follows. Consider the central motion on the plane. Let $J$ and $I_r$ be the action variables. The proof of Bertrand's theorem shows that $H$ depends on linear combination of action variables only in the Kepler $\Big(H=H(J+I_r)\Big)$ and oscillator $\Big(H=H(J+2I_r)\Big)$ cases. Now, $J\rightarrow \frac{J}{s}$ and $\varphi \rightarrow s\varphi $ is the canonical transformation relating the dynamics on the cone and on the plane. It is defined only locally due to the cyclic character of angle variable $\varphi $. However, this is irrelevant as long as the action variables are concerned. By applying this transformation we find the form of the Hamiltonian on the cone. The theorem follows.
\par
The above canonical transformation allows also to write out immediately the integral of motion for $s\not=1$ provided it is known for $s=1$. If the latter does not depend on $\varphi $ the former is globally defined. However, for $\varphi $-dependent integrals the problem is more involved. Namely, for $s$ irrational they are only locally defined; on the other hand for rational $s$ one can define the global integral by taking an appropriate function of the local one. 
\par
It is interesting to consider the quantum counterpart of Kepler and oscillator problems. One can easily show that our integrals generalize to quantum domain and their commutators define the $W$-algebras which reduce to the ones derived here in the limit $\hbar \rightarrow 0$. However, as it has been pointed out in the interesting paper \cite{b17} the situation is now much more subtle. The problem of the operator domains becomes very important. As a result the degenerate states of the Hamiltonian do not span complete representations of the symmetry algebra; we have an interesting example of formal symmetry algebra which does not determine the degeneracy. For details we refer to \cite{b17}.\\

{\bf Note added:} Just recently we have become aware of two papers dealing with Aharonov-Bohm effect in conical space \cite {b24}.\\

{\bf Acknowledgements}
We are very grateful to Cezary Gonera, Joanna Gonera and Krzysztof Andrzejewski for interesting discussion.

\end{document}